\documentclass[twocolumn,aps,prd]{revtex4}
\usepackage{graphicx,amssymb}
\usepackage[utf8]{inputenc} % Required for inputting international characters
\usepackage[normalem]{ulem}
\usepackage[urlcolor=blue,colorlinks,breaklinks=true]{hyperref}

\PassOptionsToPackage{obeyspaces,spaces,hyphens}{url}
 
 \usepackage{amsmath}
 \usepackage{comment}
 
\begin{document}
 	%My commands
 	\def\half{{1\over2}}
 	\def\shalf{\textstyle{{1\over2}}}
 	
 	\newcommand\lsim{\mathrel{\rlap{\lower4pt\hbox{\hskip1pt$\sim$}}
 			\raise1pt\hbox{$<$}}}
 	\newcommand\gsim{\mathrel{\rlap{\lower4pt\hbox{\hskip1pt$\sim$}}
 			\raise1pt\hbox{$>$}}}

\newcommand{\be}{\begin{equation}}
\newcommand{\ee}{\end{equation}}
\newcommand{\bq}{\begin{eqnarray}}
\newcommand{\eq}{\end{eqnarray}}

\newcommand{\dt}[1]{\textcolor{blue}{#1}}
\newcommand{\dr}[1]{\textcolor{red}{#1}}
\newcommand{\dg}[1]{\textcolor{green}{#1}}
\newcommand{\dm}[1]{\textcolor{magenta}{#1}}

\def\dbar{{\mathchar '26\mkern-12mu d}}
 	 	
\title{Note on bulk viscosity as an alternative to dark energy}
 	 	
\author{P. P. Avelino}
\email[Electronic address: ]{pedro.avelino@astro.up.pt}
\affiliation{Instituto de Astrof\'{\i}sica e Ci\^encias do Espa{\c c}o, Universidade do Porto, CAUP, Rua das Estrelas, PT4150-762 Porto, Portugal}
\affiliation{Departamento de F\'{\i}sica e Astronomia, Faculdade de Ci\^encias, Universidade do Porto, Rua do Campo Alegre 687, PT4169-007 Porto, Portugal}

\author{A. R. Gomes}
\email[Electronic address: ]{adalto.gomes@ufma.br}
\affiliation{Programa de P\'os-Gradua{\c c}\~ao em F\'{\i}sica, Universidade Federal do Maranh\~ao, Av. dos Portugueses, 1966, Campus Universit\'ario do Bacanga, 65085-580, S\~ao Lu\'{\i}s, Maranh\~ao, Brazil} 	 	

\author{D. A. Tamayo}
\email[Electronic address: ]{tamayo.ramirez.d.a@gmail.com}
\affiliation{Instituto de Astrof\'{\i}sica e Ci\^encias do Espa{\c c}o, Universidade do Porto, CAUP, Rua das Estrelas, PT4150-762 Porto, Portugal}

\date{\today}
\begin{abstract}

Bulk viscosity, which characterizes the irreversible dissipative resistance of a fluid to volume changes, has been proposed as a potential mechanism for explaining both early- and late-time accelerated expansion of the Universe. In this work, we investigate two distinct physical scenarios for the origin of bulk viscosity: (1) nonminimal interactions between two fluids, and (2) elastic collisions in an ideal gas. In both cases, we demonstrate that while the associated energy–momentum exchange can significantly influence fluid dynamics, overall energy–momentum conservation precludes such exchange from having any direct gravitational effect in the context of General Relativity. In case (1), we show that the standard bulk viscous energy–momentum tensor can be obtained for the two-fluid system only at the cost of the violation of all classical energy conditions: null, weak, dominant, and strong. In case (2), we consider a single fluid composed of point particles undergoing instantaneous, energy- and momentum-conserving collisions, and find that the proper pressure remains strictly non-negative, with the equation-of-state parameter confined to the interval  $[0,1/3]$. In both scenarios, achieving a sufficiently negative effective pressure to drive cosmic acceleration requires assumptions that compromise the physical viability of the model. Our results highlight some of the key physical challenges involved in modeling dark energy through bulk viscous effects.
		
\end{abstract}

\maketitle

%%%%%%%%%%%%%%%%%%%%%%%
\section{Introduction}
\label{sec:intr}

Viscosity plays a crucial role in many physical phenomena, especially in relativistic astrophysical environments characterized by extreme conditions, such as the regions around black holes or neutron stars. In these gravitationally dominated systems, it is essential to account for dissipative effects in the hydrodynamic equations. While classical descriptions of viscous fluids \cite{Landau} are sufficient for many scenarios, they become inadequate in regimes involving strong gravity or relativistic velocities. In such cases, a fully relativistic treatment of fluid dynamics is required \cite{Weinberg:1972kfs} (see also \cite{Maartens:1996vi, Andersson:2006nr}).

A natural extension of this idea has led several authors to explore the potential role of dissipative fluids in cosmology. In the context of a perfectly homogeneous and isotropic universe described by the Friedmann-Lemaître-Robertson-Walker (FLRW) metric, the spacetime symmetries imply the absence of shear viscosity and heat conduction.
For this reason, the exploration of the potential role of dissipative processes on the dynamics of the Universe has been mostly focused on relativistic bulk viscosity \cite{Maartens:1995wt, Maartens:1996vi, Zimdahl:1996ka}.

This has motivated its study as a possible driver of accelerated cosmic expansion at both early \cite{Zimdahl:1996tg, Chimento:1997ga, Mimoso:2006up, Sawyer:2006ju, Tawfik:2009mk, Bamba:2015sxa, Haro:2015ljc} and late times \cite{Fabris:2005ts, Meng:2005jy, Meng:2008dt, Avelino:2008ph, Li:2009mf, Avelino:2010pb,  Piattella:2011bs, Gagnon:2011id, Velten:2013qna, Floerchinger:2014jsa, Brevik:2014eya, Wang:2017klo, Yang:2019qza, Odintsov:2020voa, Tamayo:2020ldb}. These studies generally adopt a phenomenological framework in which a single fluid is modeled with a covariantly conserved energy–momentum tensor, corresponding to that of a perfect fluid with an additional term a proper pressure contribution associated with bulk viscosity.  While such phenomenological approaches may be sufficient for an effective description within classical fluid dynamics, they fail to provide a comprehensive and fully consistent physical framework for the underlying mechanisms responsible for the emergence of bulk viscosity.

In Eckart’s theory \cite{Eckart:1940te}, bulk viscous pressure is treated as a small perturbation to the equilibrium pressure. The Israel–Stewart theory \cite{IsraelStewart1979} avoids the acausal propagation of perturbations inherent to Eckart’s formulation, but it remains perturbative despite incorporating higher-order terms. Regardless of these limitations, any significant cosmological impact of bulk viscosity should ultimately arise from far-from-equilibrium physics. Indeed, if bulk viscosity is proposed as an explanation for dark energy, its contribution must dominate the total pressure. This requirement renders a perturbative description of bulk viscosity inadequate for such a role. Thus, in contrast to the standard assumption that the fluid’s true energy–momentum tensor is that of a perfect fluid supplemented by an ad hoc bulk-viscous term, here we adopt a different perspective. Specifically, we consider two distinct physical scenarios for the origin of bulk viscosity: (i) nonminimal interactions between two different fluids, and (ii) elastic collisions in an ideal gas. We then examine the potential cosmological implications of these scenarios, with particular emphasis on their viability as potential dark energy mechanisms.

Previous studies \cite{Li:2009mf, Velten2011, Giovannini2015, Barbosa:2017} have already identified problems with the hypothesis that bulk pressure alone can drive the late-time acceleration of the Universe, suggesting that an additional dark energy component remains necessary. In most cases, these issues arise at the perturbative level rather than in the background evolution. In contrast, this study explores more fundamental physical limitations of attributing late-time cosmic acceleration solely to a bulk viscous fluid.

The outline of this paper is as follows. In Sec. II, we consider nonrelativistic bulk viscosity and demonstrate that its impact on the dynamics of the Universe is negligible. In Sec. III, we explore a relativistic scenario where bulk viscosity arises from nonminimal interactions between two fluids, critically examining its potential effects on the accelerated expansion of the Universe. In Sec. IV, we analyze the case of an idealized single-fluid model composed of point particles interacting microscopically conserving energy and momentum. Finally, our conclusions are presented in Sec. V.

Throughout this work, we adopt the metric signature $[-,+,+,+]$ and use natural units where $c = 1$.

%%%%%%%%%%%%%%%%%%%%%%%%%%%%%%%%%%%%%%%%%%%
\section{Nonrelativistic bulk viscosity}
%%%%%%%%%%%%%%%%%%%%%%%%%%%%%%%%%%%%%%%%%%%

Bulk viscosity characterizes the irreversible response of a fluid to changes in volume and may be associated with an effective pressure experienced by the fluid that differs from its true proper pressure. Here, we consider a fluid without shear viscosity or heat flux, composed of stable, nonrelativistic particles, and assume that this pressure difference corresponds to the bulk viscous pressure:
\be
\Pi = p_{\rm eff}-p = - \zeta \nabla \cdot \vec v \, , \label{Pi} 
\ee
where $\nabla \cdot \vec v$ is the divergence of the fluid's 3-velocity field, and $\zeta > 0$ is the bulk viscosity coefficient. This coefficient generally depends on the fluid’s microscopic properties, which may vary with time and position.

In a nonrelativistic regime, the internal energy of the fluid is approximately conserved, which implies that the continuity equation
\be
\frac{\partial \rho}{\partial t} + \nabla \cdot (\rho \vec v) = \frac{d \rho}{d t} +\rho \,  \nabla \cdot \vec v = 0 \label{continuity}
\ee
holds to first order in $|\vec v|$. When combined with Eq.~\eqref{Pi}, this yields
\be
\Pi=\zeta \frac{d \ln \rho}{dt}\,. \label{Pirho}
\ee

The energy density of a homogeneous volume element $V$ containing a fixed number $N$ of nonrelativistic particles of constant proper mass $m$ is approximately given by
\be
\rho=\frac{m N}{V}\,. \label{rhonr}
\ee
Combining Eqs.~(\ref{Pirho}) and (\ref{rhonr}) we obtain an approximate relation between the bulk viscous pressure and the volume change
\be
\Pi=-\zeta \frac{1}{V}\frac{d V}{dt} \,.
\ee
The work done by the bulk viscous pressure in an expanding system is then given approximately by
\be
W_\Pi= \int_{V_i}^{V_f} \Pi dV = - \int_{V_i}^{V_f}  \zeta \frac{dV}{dt} \frac{dV}{V} < 0\,.
\ee
The negative sign indicates that the work done by the viscous pressure opposes the volume change.

In the absence of heat transfer, ${\dbar Q}=0$, the internal energy $U$ of the fluid evolves according to
\be
dU + p_{\rm eff} dV = 0\,.
\ee
Substituting  $U=\rho V$, one obtains
\be
d\rho + (\rho + p_{\rm eff}) \frac{dV}{V} = 0\,.
\label{rho}
\ee
In a nonrelativistic regime with $|p_{\rm eff}| \ll \rho$ Eq. (\ref{rho}) implies that the internal energy $U$ of the fluid is approximately conserved [consistently with Eq.~(\ref{continuity})].

\subsection{Application to cosmology}

In a perfectly homogeneous and isotropic Friedmann-Lemaître-Robertson-Walker universe, one can define a local inertial frame in which the velocity of nearby fluid elements follows the Hubble law:
\be
\vec v = H \vec r \,,
\ee
where $H=d \ln a/dt$ is the Hubble parameter and $a(t)$ is the scale factor. Using the fact that $\nabla \cdot \vec r =3$, it follows that the bulk viscous pressure in a cosmological setting takes the form
\be
\Pi=- \zeta \nabla \cdot \vec v = - 3 \zeta  H \,. \label{PiH}
\ee
Thus, in cosmology, bulk viscosity manifests as an effective pressure term proportional to the Hubble parameter. Substituting Eq. \eqref{PiH} into Eq. \eqref{rho}, and using Eq. \eqref{Pi}  along with the relation $V \propto a^3$, the continuity equation becomes
\be
\frac{d\rho}{dt} +3 H (\rho + p)  =  9 H^2 \zeta\,.
\ee
In a nonrelativistic regime with $|p_{\rm eff}|=|p-3\zeta H| \ll \rho$ the evolution of the energy density of the fluid with the scale factor is given approximately by $\rho \propto a^{-3}$.

\section{Relativistic bulk viscosity: two nonminimally interacting fluids}

Here, we shall assume that the energy content of the Universe is composed of two nonminimally interacting fluids. For the sake of definiteness, we shall assume that the primary fluid is a perfect fluid characterized by an energy-momentum tensor $\bf T$ with components given by
\be
T^{\mu \nu} = (\rho + p) U^\mu U^\nu + p g^{\mu \nu} \,,\label{Tperfect}
\ee
where $U^\mu$ are the components of the 4-velocity $\bf U$ of the fluid, $g^{\mu \nu}$ are the components of the metric tensor $\bf g$, $\rho$ and $p$ are, respectively, its proper energy-density and pressure, defined by
\bq
\rho&=&T^{\mu \nu} U_\mu U_\nu \,,\\
p &=& \frac13  h_{\mu \nu} T^{\mu \nu}\,,
\eq
$U_\mu =g_{\mu \alpha} U^\alpha$, $g^{\mu \alpha}g_{\alpha \nu}=\delta^\mu_\nu$ where $\delta^\mu_\nu$ is the Kronecker delta function, and
\be
h_{\mu \nu} = g_{\mu \nu} + U_\mu U_\nu\,. \label{hmunu}
\ee

Let us also assume that this perfect fluid is subject to a bulk viscous pressure given by
\be
\Pi = -\zeta \nabla_\mu U^\mu \,,
\ee
as a result of the interaction with a secondary fluid whose energy–momentum tensor is given by ${\bf T}_*$. In this case, the energy and momentum of the two fluids are generally not conserved separately. Instead, energy and momentum may be exchanged between the fluids. This exchange is described by the equations
\bq
\nabla_\nu T^{\mu \nu} &=&  -\nabla_\nu ( \Pi h^{\mu \nu}) \,, \\ \nabla_\nu T_*^{\mu \nu} &=& \nabla_\nu ( \Pi h^{\mu \nu}) \,,
\eq
where
\be
h^{\mu \nu} = g^{\mu \alpha}g^{\nu \beta} h_{\alpha \beta}\,,
\ee
with the two nonminimally interacting fluids interacting fluids exchanging energy and momentum via the source term $Q^\mu = \nabla_\nu(\Pi h^{\mu \nu})$.

One may then define effective energy–momentum tensors
\bq
T^{\mu \nu [\rm eff]} &=& T^{\mu \nu} + \Pi h^{\mu \nu} \,, \\ \label{Teffprimary}
T_*^{\mu \nu [\rm eff]} &=& T_*^{\mu \nu} - \Pi h^{\mu \nu} \,,
\eq
which are covariantly conserved ($\nabla_\nu T^{\mu \nu [\rm eff]}=0$ and $\nabla_\nu T_*^{\mu \nu [\rm eff]}=0$). The full energy–momentum tensor appearing on the right-hand side of the Einstein equations is the sum of the energy–momentum tensors of both fluids, thus implying that
\be
T^{\mu \nu [\rm full]}=T^{\mu \nu}+T_*^{\mu \nu}=T^{\mu \nu [\rm eff]}+T_*^{\mu \nu [\rm eff]}\,.
\ee
$T^{\mu \nu [\rm full]}$ is in general different from the primary fluid effective energy–momentum tensor $T^{\mu \nu [\rm eff]}$. The only exception occurs if $T_*^{\mu \nu [\rm eff]} = T_*^{\mu \nu} - \Pi h^{\mu \nu}=0$ or, equivalently, if $T_*^{\mu \nu} = \Pi h^{\mu \nu}$. This particular case formally corresponds to the standard bulk viscous fluid that appears in the literature: a single perfect fluid whose energy–momentum tensor is that of a perfect fluid with an added bulk viscosity term. However, this scenario would not only require that the 4-velocities of both fluids be equal at every space-time point (${\bf U}_*={\bf U}$), but also a vanishing proper energy density of the secondary fluid:
\be
\rho_*=T_*^{\mu \nu} U_{*\mu} U_{*\nu} = T_*^{\mu \nu} U_{\mu} U_{\nu}  = 0\,,
\ee
where we have taken into account that $U^\mu U_\mu=-1$. Moreover, taking into account that $g_{\mu \nu}  g^{\mu \nu} = 4$, one finds that the proper pressure of the secondary fluid would need to be equal to the bulk pressure:
\be
p_* = \frac13  h_{*\mu \nu} T^{\mu \nu}_* =\frac13  h_{\mu \nu} T^{\mu \nu}_*=\Pi  \,,
\ee
where ${\bf h}_{*}$ is defined as in Eq. (\ref{hmunu}) but with $\bf U$ replaced by $\bf U_*$. However, having $\rho_*=0$ and $p_*=\Pi < 0$ would violate the null, weak, dominant and strong energy conditions.

\subsection{Homogeneous and isotropic fluids}

Consider the particular case of two homogeneous and isotropic fluids in a flat, homogeneous, and isotropic universe, described by the Friedmann-Lemaître-Robertson-Walker metric, whose line element can be written as
\be
ds^2=-dt^2  +a^2(t) d \vec q \cdot d \vec q \,,
\ee
where $t$ is the cosmic time, $\vec q$ are comoving Cartesian coordinates. In this case, $U^0=1$ and $U^i=0$, which implies that the bulk viscous pressure is given by
\be
\Pi = -\zeta \nabla_\mu U^\mu = - 3 \zeta H \,.
\ee

The evolution of the energy densities of the two fluids then obeys
\bq
\frac{d \rho}{dt}+3H(1+w)\rho &=& -3 H \Pi = 9 \zeta H^2\,,\\
\frac{d \rho_*}{dt}+3H(1+w_*)\rho_* &=& 3 H \Pi = -9 \zeta H^2 \,,
\eq
where $w = p/\rho$ and $w_*=p_*/\rho_*$ are the equation-of-state parameters of the primary and secondary fluids, respectively. These two equations can also be written as
\bq
\frac{d \rho}{dt}+3H(1+w^{[\rm eff]}) \rho &=& 0 \,,\\
\frac{d \rho_*}{dt}+3H(1+w_*^{[\rm eff]})\rho_* &=& 0\,.
\eq
where $w^{[\rm eff]} = (p+\Pi)/\rho$ and $w_*^{[\rm eff]}=(p_*-\Pi)/\rho_*$ are the effective equation of state parameters of the primary and secondary fluids, respectively. 

The Raychaudhuri equation is given by
\be
\frac{\ddot a}{a}= - \frac{4\pi G}{3} \rho^{[\rm full]} \left(1+3w^{[\rm full]}  \right) \,,
\ee
where, $\rho^{[\rm full]}=\rho+\rho_*$ and
\be
w^{[\rm full]} = \frac{w^{[\rm eff]}\rho + w_*^{[\rm eff]} \rho_*}{\rho^{[\rm full]}} \label{wfull}
\ee
is different from $w^{[\rm eff]} $, except when i) $\rho_*=0$ or ii) $w^{[\rm full]}=w^{[\rm eff]}  =w_*^{[\rm eff]}$. Let us consider separately these two possibilities.

If $H \neq 0$, the condition i) $\rho_*=0$ implies that $p_* = -3 \zeta H = \Pi$, and the total pressure becomes $p^{[\rm full]}=p+p_*= p -3 \zeta H = p + \Pi = p^{[\rm eff]}$, while the total density would be equal to $\rho^{[\rm full]}=\rho+\rho_*=\rho$.  
However, this scenario should be discarded because the secondary fluid with $\rho_*=0$ and $p_* = \Pi < 0$ would imply a violation of the null, weak, dominant and strong energy conditions (see \cite{Kontou:2020bta} for a recent review of energy conditions in general relativity and quantum field theory). Condition ii) $w^{[\rm full]}=w^{[\rm eff]}  =w_*^{[\rm eff]}$ implies that
$w=w_*-{\Pi}(\rho^{-1} + \rho_*^{-1})$. In that case, assuming that $\Pi < 0$, one has $w > w^{[\rm full]} > w_*$.

More generally, the cosmic acceleration condition $\ddot{a} > 0$ implies that
\bq
w^{[\rm full]} &=& \frac{p+p_*}{\rho^{[\rm full]}} \nonumber \\
&=& \frac{\rho}{\rho^{[\rm full]}} w + \left(1-\frac{\rho}{\rho^{[\rm full]}} \right) w_*< -\frac{1}{3}\,.
\eq
This, in turn, implies that $w < -1/3$ or $w_* < -1/3$. Hence, at least one of the fluids must violate the strong energy condition in order for the universe to accelerate, regardless of the role played by bulk viscosity.

The issues of causality and stability (such as superluminal propagation and unstable equilibrium states present in the full relativistic Eckart first-order formalism) can be avoided by employing the Israel–Stewart second-order formalism. However, our primary goal in this section was to illustrate, within the simplest possible framework, that although the energy–momentum exchange between the fluids can significantly affect their individual dynamics, overall energy–momentum conservation prevents such exchange from producing any direct gravitational effect in the context of General Relativity. This conclusion should remain valid irrespective of the formalism adopted.

\section{Relativistic bulk viscosity: single fluid}

The assumption commonly made when discussing the cosmological impact of bulk viscosity is that of a single perfect fluid with a covariantly conserved energy–momentum tensor $\bf T$ equal to the effective energy-momentum tensor ${\bf T}^{[\rm eff]}$ of the primary fluid defined in the previous section [Eq. \eqref{Teffprimary}] or, equivalently, with
\be \label{single bulk viscous fluid}
T^{\mu \nu} = {\tilde T}^{\mu \nu} - \zeta\nabla _\alpha U^\alpha (g^{\mu \nu}+ U^\mu U^\nu)\,,
\ee
where 
\be
 {\tilde T}^{\mu \nu} = (\tilde \rho + \tilde p) U^\mu U^\nu + \tilde p g^{\mu \nu} \, ,
\ee
so that $\rho = \tilde \rho$ and $p = \tilde p - \zeta  \nabla _\alpha U^\alpha$. In this case, a violation of the strong energy condition is still required for the fluid to account for the acceleration of the Universe. However, the null, weak, and dominant energy conditions may or may not be violated, depending on the magnitude of the bulk viscous pressure.

In the following discussion, we shall therefore assume a single fluid, in which case the emergence of the bulk viscosity term must result from nonminimal interactions within the fluid itself. Here, we assume that this fluid can be approximated as a collection of stable point particles with fixed mass, subject to instantaneous energy and momentum conserving interactions. Under these assumptions, we demonstrate that the pressure of such a fluid can never be negative.

The trace of the energy–momentum tensor of a fluid element containing $N$ particles, small enough for its proper pressure and density to be approximately constant, is given by (see, for example, \cite{Weinberg:1972kfs}):
\be
{T^{\mu}}_{\mu}  = -\rho + 3p = - \sum_{i=1}^N \frac{m_i}{\gamma_i} \delta^3 (\vec r - \vec r_i)\,, \label{Tpparticles}
\ee
where $p = \frac{1}{3}({T^x}_x + {T^y}_y + {T^z}_z)$ is the isotropic pressure, $m_i$ is the proper mass, $v_i$ is the speed, and $\vec r_i$ is the position of the $i$th particle, $\gamma_i = (1 - v_i^2)^{-1/2}$ is the corresponding Lorentz factor, and $\delta^3(\vec r)$ is the three-dimensional Dirac delta function. All quantities are defined in a local inertial frame momentarily comoving with the fluid element.
Integrating over the fluid element volume, one finds
\be
\int {T^{\mu}}_{\mu} d V = -\int (\rho-3p) dV = - \sum_{i=1}^N \frac{m_i}{\gamma_i} \,,
\ee
or, equivalently,
\bq
\int p dV &=& \frac13 \left(\int \rho dV   - \sum_{j=1}^N \frac{m_j}{\gamma_j}\right) \nonumber\\
&=& \frac13 \left( \sum_{i=1}^N m_i \gamma_i - \sum_{j=1}^N \frac{m_j}{\gamma_j}\right) \ge 0\,.
\eq
On the other hand,
\bq
\int p dV &\le& \frac13\int \rho dV\,.
\eq
Hence, the pressure associated to a fluid constituted exclusively of point particles with fixed proper mass is constrained to lie within the interval $0 \le p \le \rho/3$. The lower bound corresponds to dust, where $\gamma_i=0$ (or, equivalently, $v_i=0$) for all $i=1,\dots,N$, while the upper bound is approached in the ultrarelativistic limit, $\gamma_i \to \infty$.  

Although energy–momentum exchange between particles can significantly 
affect the dynamics of the fluid, overall energy–momentum conservation in General Relativity prevents such interactions from having a direct gravitational impact. Moreover, due to the resulting constraint on the fluid’s equation of state, such a fluid cannot be the source of the observed acceleration of the Universe.  This conclusion is independent of whether or not the fluid is in thermodynamic equilibrium.

In \cite{Gagnon:2011id} the authors propose a microscopic model for dark energy based on a fluid composed of self-interacting spin-zero particles, in which the dependence of the bulk-viscosity coefficient $\zeta$ on the scalar temperature is derived under the assumptions of near equilibrium and weak coupling. Within this framework, they show that at sufficiently late times (or, equivalently, at low scalar temperatures), the bulk viscosity becomes the dominant contribution to the total pressure of the fluid, leading to a dark energy equation of state. Still, the validity of the near-equilibrium approximation in the regime where bulk viscosity dominates may warrant further investigation. In contrast, our single-fluid model makes only the assumption that the cosmic medium can be effectively described as a collection of stable, pointlike particles of fixed mass, undergoing instantaneous, energy- and momentum-conserving interactions. This provides a minimal microscopic picture underlying the effective microscopic dynamics, without invoking any specific field-theoretic realization of the constituents. Note, however, that even though a kinetic-theory description provides an excellent description of many physically relevant fluids, it does not apply universally — certain media, such as scalar-field condensates, cannot be described in terms of such pointlike particle interactions.

\section{Conclusions}\label{sec:conc}

In this work, we presented a nonstandard perspective on bulk viscosity, interpreting it as a dynamical addition to the fluid’s effective pressure arising from nonminimal interactions, rather than a (small) perturbation to the equilibrium pressure. We investigated two different physical scenarios for its origin: one involving two nonminimally interacting fluids, and another based on nonminimal interactions among the constituents of an ideal gas. In both cases, we identified fundamental issues that prevent these models from 
consistently accounting for the observed accelerated expansion of the Universe.

In the case of two nonminimally interacting fluids, we showed that achieving accelerated expansion requires at least one of the fluids to have an equation of state that violates the strong energy condition. From this perspective, the presence of bulk viscous pressure does not eliminate the need for dark energy. This effectively reduces the setup to a dark energy model with an expansion-dependent proper pressure, thereby undermining the original motivation for invoking bulk viscosity as an explanation for the observed cosmic acceleration. Importantly, we found that the commonly used phenomenological model in the literature, when framed in this framework, corresponds to an unphysical configuration in which the secondary fluid has vanishing energy density but nonzero negative pressure, leading to violations of the null, weak, dominant, and strong energy conditions.

Moreover, we examined the scenario where bulk viscosity arises from nonminimal interactions among the constituents of a single fluid composed of point particles. We have shown that the pressure is bounded by $0 \le p \le \rho/3$, thereby preventing the emergence of the negative pressure necessary to drive cosmic acceleration. In fact, the energy–momentum exchange, whether between fluids or among constituents, does not have a direct impact on the true energy–momentum tensor that sources the Einstein field equations. 
Instead, any potential influence on cosmic acceleration must result from the dynamical modifications that these interactions induce in the behavior of the fluid or its constituents.

Although both of these models fail to establish a direct link between bulk viscosity and the negative pressure required to sustain an accelerated expansion, they highlight the theoretical challenges involved in modeling dissipative processes in a gravitational context and reveal significant physical limitations that any viable model must address. Given these challenges and the idealized nature of the models considered, claims that bulk viscosity can drive the accelerated expansion of the Universe should be approached with careful scrutiny.

While obtained under specific assumptions, our results suggest that any potential role of bulk viscosity in cosmic acceleration would likely require intricate microphysical mechanisms beyond those considered here, or the inclusion of extensions to general relativity. Exploring such directions may yield a fully consistent framework in which bulk viscosity provides a more robust explanation for the late-time acceleration of the universe.

%%%%%%%%%%%%%%%%%%%%%%%%%%%%%%%%%%%%%%%%%%%%%%%%%%%%%
\begin{acknowledgments}

We thank our colleagues of the Cosmology group at Instituto de Astrofísica e Ciências do Espaço for enlightening discussions. P.P.A. and D.A.T. acknowledge the support by Fundação para a Ciência e a Tecnologia (FCT) through the
research Grant No. UID/04434/2025. A.R.G. acknowledge the support by Coordena\c c\~ao de Aperfei\c coamento de Pessoal de N\'ivel Superior - Brasil (CAPES) - Finance Code 001, CNPq through the research Grant No. 402830/2023 and  Fundação de Amparo à Pesquisa e ao Desenvolvimento Científico e Tecnológico do Maranhão (FAPEMA) through the research Grants No. COOPI-07838/17 and No. PV3-06550/18.
\end{acknowledgments}
%%%%%%%%%%%%%%%%%%%%%%%%%%%%%%%%%%%%%%%%%%%%%%%%%%%%%%%%%%
 
\bibliography{Bviscosity}
 	
 \end{document}